\documentclass[twocolumn,showpacs,preprintnumbers,amsmath,amssymb,aps]{revtex4}
\usepackage{graphicx}

\begin{document}

\preprint{Na on graphene}

\title{The stability of graphene band structures against an external periodic perturbation ; Na on Graphene}
\author{C. G. Hwang$^{1,3}$}
\author{S. Y. Shin$^1$}
\author{Seon-Myeong Choi$^1$}
\author{N. D. Kim$^{1,4}$}
\author{S. H. Uhm$^1$}
\author{H. S. Kim$^1$}
\author{C. C. Hwang$^2$}
\author{D.Y. Noh}
\author{Seung-Hoon Jhi$^1$}
\author{J. W. Chung$^1$} \email{jwc@postech.ac.kr}

\affiliation{$^1$Department of Physics, Pohang University of Science and Technology, Pohang, Korea}
\affiliation{$^2$Beamline Research Division, Pohang Accelerator Laboratory (PAL), Pohang, Korea}
\affiliation{$^3$Present address: Materials Sciences Division, Lawrence Berkeley National Laboratory, Berkeley, CA,
USA} \affiliation{$^4$Present address: Department of Physics, Columbia University, New York, NY, USA}
\affiliation{$^5$Department of Materials Science and Engineering, Gwangju Institute of Science and Technology, Gwangju,
Korea}

\date{\today}

\begin{abstract}
We report that the $\pi$ band of graphene sensitively changes as a function of an external potential induced by Na
especially when the potential becomes periodic at low temperature. We have measured the band structures from the
graphene layers formed on the 6H-SiC(0001) substrate using angle-resolved photoemission spectroscopy with synchrotron
photons. With increasing Na dose, the $\pi$ band appears to be quickly diffused into background at 85 K whereas it
becomes significantly enhanced its spectral intensity at room temperature (RT). A new parabolic band centered at
$k\sim$1.15 \AA$^{-1}$ also forms near Fermi energy with Na at 85 K while no such a band observed at RT. Such changes
in the band structure are found to be reversible with temperature. Analysis based on our first principles calculations
suggests that the changes of the $\pi$ band of graphene be mainly driven by the Na-induced potential especially at low
temperature where the potential becomes periodic due to the crystallized Na overlayer. The new parabolic band turns to
be the $\pi$ band of the underlying buffer layer partially filled by the charge transfer from Na adatoms. The five
orders of magnitude increased hopping rate of Na adatoms at RT preventing such a charge transfer explains the absence
of the new band at RT.
\end{abstract}


\maketitle

\section{Introduction}

Recent spur on the study of graphene has been motivated mostly by the unusual nature of its effectively massless charge
carriers so-called Dirac fermions leading to some exotic electronic properties, its potential application for future
nanometer-scale devices, on the other hand, has also been under extensive exploration. Graphene is a flat atomic single
layer of graphite having a hexagonal crystalline symmetry with two atoms per unit cell. Its low energy band structure
shows the almost linear $\pi$ bands with a vanishing density of states at the Dirac point of charge neutrality where
the valence band ($\pi$ band) and conduction band ($\pi^*$) meet at the K point of the hexagonal Brillouin zone. With
increasing reports on the electronic properties of graphene layers including the anomalous quantum Hall
effects\cite{Zhang05,Kane05,Li07}, the possibility of utilizing such features for atomic-scale devices seems to
increase as seen also in the bilayer graphene \cite{Ohta07}. Ohta et al. show that the adsorption of Potassium on
bilayer graphene modifies the potential symmetry in the graphene layers resulting in the change of energy gap at the K
point. This provides a way of controlling the conductivity of the charge carriers for the switching capability in the
atomic scale. Moreover recent calculations also provide a way to control the electrical properties of graphene by
utilizing the chiral nature of the Dirac fermions of graphene.\cite{Park08} Park et al., show that the Dirac fermions
of graphene can be quite sensitive to an external potential resulting, for example, the anisotropic renormalization of
group velocity at the K point driven essentially by the chiral character of charge carriers of graphene. This feature
becomes more prominent when the potential becomes periodic with a periodicity L comparable to that of the graphene
lattice constant a$_0\sim$1.42 \AA. In addition, the life-time of Dirac fermions becomes significantly reduced when
L$\ll$a$_0$ by enhancing back-scatterings from the potential between states with momentums of $+k$ and $-k$
\cite{Ando1,Ando2}. No experimental study on the stability of graphene bands against a foreign perturbation, however,
has been made especially for a potential with L$\sim$a$_0$ despite several earlier angle-resolved photoemission
spectroscopy (ARPES) studies\cite{Reinert,Lanzara,Gweon}.

We report here an example of such a study that the presence of an external potential on graphene significantly affects
the nature of Dirac fermions not only by the periodicity of the potential but also more remarkably by the degree of
order of the potential especially when L$\sim$a$_0$. We have investigated the behavior of the $\pi$ band along the
$\Gamma-K$ direction as we adsorb Na on graphene at two different temperatures, 85 K and room temperature (RT). We find
remarkable changes in the $\pi$ band as a function of Na dose at both temperatures. We also notice the appearance of a
parabolic band centered at k=1.15 \AA$^{-1}$ only at 85K. Furthermore the $\pi$ band exhibits quite different behavior
with Na on a bi-layer graphene from that of a single layer graphene. We mainly discuss such changes in the $\pi$ band
of graphene in terms of the variation of the induced potential by Na adsorption. We also discuss the origin and
characteristics of the new parabolic band from our band calculation.

\section{Methods}

We have obtained our ARPES data by using the ARPES chamber at the beamline 3A2 of the Pohang Accelerator Laboratory in
Korea using synchrotron photons of energy 34 eV. The chamber was equipped with a hemispherical Scienta R4000 electron
analyzer which provides an overall energy resolution of 113 meV. All measurements were made with the chamber under
1$\times$10$^{-10}$ Torr. An n-type 6H-SiC(0001) substrate was pre-treated by annealing at 900 $^{\circ}$C under the Si
flux generated from a resistively heated Si wafer at 1150 $^{\circ}$C to obtain the $\sqrt{3}\times\sqrt{3}$ phase. We
then continued to anneal the substrate further to obtain a carbon rich surface with the $6\sqrt{3}\times6\sqrt{3}$
phase as reported earlier.\cite{**} Single layer and bilayer graphenes were self-assembled when this precursor phase
was annealed for an extended time period. We have used a commercial (SAES) getter source to deposit Na on the graphene
surface and measured work function change to estimate Na coverage.

In order to interpret our band data, we have also performed {\em ab initio} pseudopotential total energy calculations
with a plane-wave basis set.\cite{Cohen} The exchange-correlation of electrons was treated within the generalized
gradient approximation (GGA) as implemented by Perdew-Berke-Enzelhof.\cite{Perdew} The cutoff energy for the expansion
of wave-functions and potentials in the plane-wave basis was chosen to be 400 eV, and the Brillouin zone sampling was
done with the Monkhorst-Pack special k-point method with a grid of 5$\times$5$\times$1 for the $\sqrt{3}\times\sqrt{3}$
buffer layer phase. We used the projector augmented wave pseudopotentials as provided by the software package (Vienna
Ab-initio Simulation Package).\cite{Kresse} The atomic relaxation was carried out until the Helmann-Feynman forces were
less than 0.02 eV/\AA. The vacuum layer in the supercell used in our calculations is set to be 8 \AA, which is enough
to minimize the artificial interlayer interaction.

\section{Results and Discussions}

\begin{figure}[t]
\includegraphics{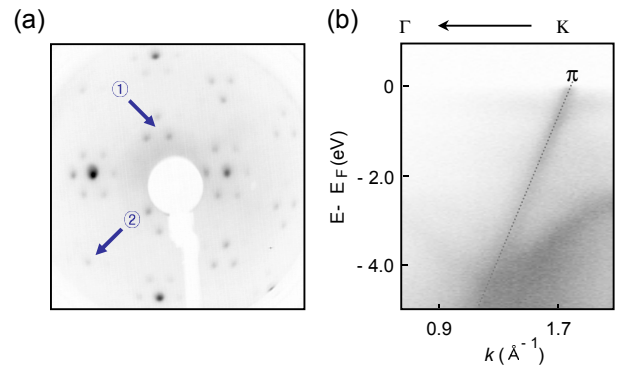}
\caption{\label{fig:fig1} (Color online) (a) A typical LEED pattern from a single layer graphene formed on SiC
substrate measured with a primary beam energy of 98 eV. The characteristic spots (arrows 1 and 2) from the single layer
graphene coexist with the spots (arrow 1) from the $\sqrt{3}\times\sqrt{3}$ phase. (b) The formation of a single layer
graphene is finally confirmed by observing the unsplit linear $\pi$ band of graphene in the ARPES intensity map near
the K point. The band was measured with synchrotron photons of energy 34 eV along the $\Gamma-K$ direction.}
\end{figure}

The phase of single layer graphene formed on SiC(0001) surface was first indicated by its characteristic LEED spots as
seen in Fig.~\ref{fig:fig1}(a). While weakening the spots from the underlying $\sqrt{3} \times \sqrt{3}$ phase (arrow
1), the LEED spots unique to the graphene $1 \times 1$ phase (arrow 2) appear with growing intensity as annealing
continued. A primary electron energy of 98 eV is used to monitor the changes in the LEED pattern during the preparation
process. The electronic structure of the graphene thus prepared has been measured using ARPES along the $\Gamma-K$
direction of graphene up to the second Brillouin zone of the $\sqrt{3}\times\sqrt{3}$ phase [see
Fig.~\ref{fig:fig5}(b)]. Photoelectron intensity map near the zone boundary $K$ shows the well known graphene $\pi$
band [Fig.~\ref{fig:fig1}(b)] an indicative of a single layer graphene. This linear dispersion of the $\pi$ band
suggests that the charge carriers in graphene can be considered as massless Dirac fermions. The single layer graphene
suggested by LEED, however, has been finally confirmed by observing the unsplit linear $\pi$ band along the $\Gamma-K$
direction as shown in Fig.~\ref{fig:fig1}(b) before developing into the split $\pi$ band with further annealing due to
the formation of multiple graphene layeres.\cite{Ohta07} Here we observe only one band rather than the two forming a
conical shape $\pi$ band due to the chiral property of graphene.\cite{Shir95,Mucha07} From the slope of the band we
estimate the Fermi velocity $v_F$=1.1$\pm$0.074$\times$10$^6$ m/s in good agreement with earlier value of 10$^6$ m/s
from Shubnikov-de Haas oscillations\cite{Novo05} where the effective carrier mass of graphene was estimated to be about
0.02 $\sim$ 0.07 $m_0$ with $m_0$ being the free electron mass.\cite{Novo05}

\begin{figure*}[t]
\includegraphics{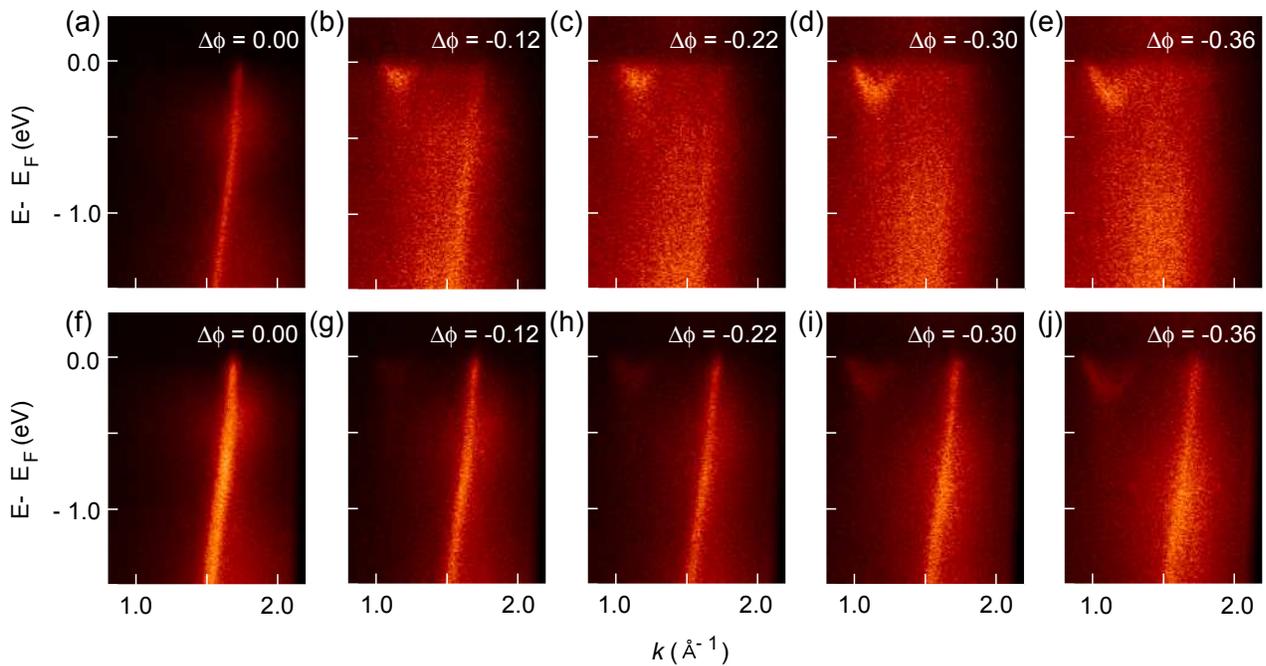}
\caption{\label{fig:fig2} (Color online) (a)-(e) The progressive change of the band structure of a single layer
graphene along the $\Gamma-K$ direction as a function of Na dose at 85 K. With increasing Na dose, the linear $\pi$
band quickly deteriorates both in intensity and in band width while a new parabolic band develops. (f)-(j)
Corresponding change for a bi-layer graphene at 85 K.}
\end{figure*}

In Fig.~\ref{fig:fig2}, we show progressive changes of the $\pi$ band with increasing function of Na dose on a single
layer (a-e) and a bi-layer (f-j) graphene at temperature 85 K. The amount of Na atoms adsorbed is indicated by work
function change ($\Delta\phi$) induced by Na adsorption. The line-width of the $\pi$ band in (f) appears to be much
broader than that in (a) of the single layer graphene. Although our resolution does not resolve the broadened band into
two or more split bands, the measured line-width of 0.08$\pm$0.02 \AA$^{-1}$ indicates the $\pi$ band from a bi-layer
graphene.\cite{Ohta07} From (a-e), one immediately notices that the adsorption of Na on the single layer graphene
quickly deteriorates the $\pi$ band by significantly weakening its spectral intensity as well as broadening the
line-width, and concomitantly accompanies a new parabolic band near Fermi energy. The deterioration of the $\pi$ band
becomes worse while the parabolic band becomes stronger with increasing Na coverage. The corresponding changes induced
by Na adsorption on the bi-layer graphene (f-j) are quite different from those on the single layer graphene; The $\pi$
band remains strong in intensity while slightly broadening with increasing Na dose. The parabolic band is still there
but becomes much weaker compared to the one on the single layer graphene at the same dose of Na.

\begin{figure}[b]
\includegraphics{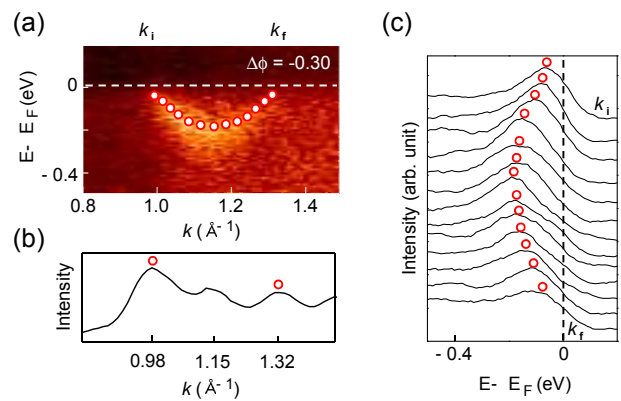} \caption{\label{fig:fig3} (Color online) (a) Photoemission intensity map of a new band from a
single layer graphene along the $\Gamma-K$ direction at Na dose of $\Delta\phi$=$-$0.30. (b) The new band appears to be
a parabolic shape with a minimum at $\sim$1.15 \AA$^{-1}$~, which is apparent in the MDC at the Fermi level and also in
(c) where EDCs are shown between $k_i$=0.98 \AA$^{-1}$ and $k_f$=1.32 \AA$^{-1}$. The red circles denote the maximua in
intensity.}
\end{figure}

We first investigate the nature of the parabolic band. Figure~\ref{fig:fig3} shows this new band at $\Delta\phi$=0.30
eV. Red circles in Fig.~\ref{fig:fig3}(a) are the maxima in spectral intensity obtained by fitting the energy
distribution curves (EDCs) with a Lorentzian function for 0.98 \AA$^{-1}\leq$k$\leq$1.32 \AA$^{-1}$ [see
Fig.~\ref{fig:fig3}(c)]. The dispersion of the new band estimated from the momemtum distribution curve at Fermi energy
in Fig.~\ref{fig:fig3}(b) appears to be parabolic with a minimum at $k\sim$1.15 \AA$^{-1}$~ for E=0.19 eV below the
Fermi level. We obtain charge carrier mass m=0.58$\pm$0.07 $m_0$ and $v_F$=3.4$\pm$0.41$\times$10$^5$ m/s, from the
parabolic dispersion of the new band. The effective mass of charge carriers of this band is 8.3$\sim$29 times heavier
than that of the Dirac electrons of the clean graphene without Na.

\begin{figure}[t]
\includegraphics{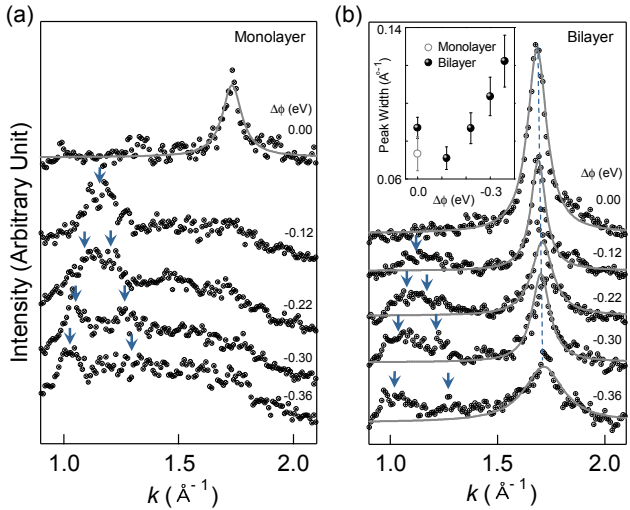}
\caption{\label{fig:fig4} (Color online) The spectral changes of the MDCs with increasing Na dose measured in terms of
$\Delta\phi$ obtained from (a) single layer graphene and (b) bi-layer graphene. The down arrows indicate the intensity
maxima of the new parabolic band. The solid gray curves are Lorentzian fit-curves of the $\pi$ band of graphene. The
vertical dotted line in (b) shows the shift of the $\pi$ band toward higher binding energy with increasing Na dose.}
\end{figure}

\begin{figure}[t]
\includegraphics{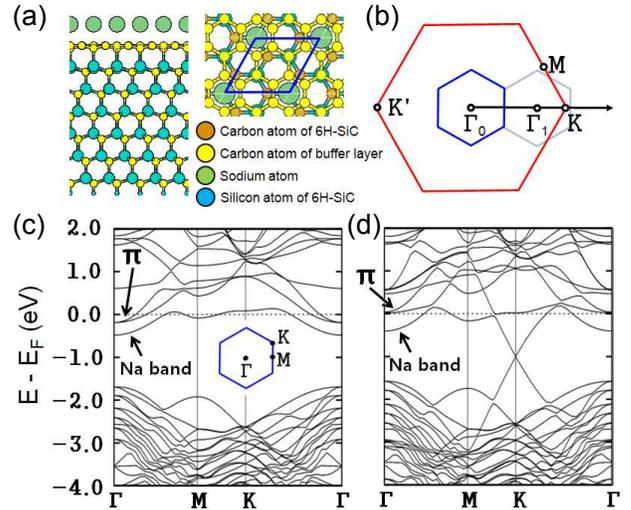}
\caption{\label{fig:fig5} (Color online) (a) Side (left) and top (right) view of the atomic arrangement for the
$\sqrt{3}\times\sqrt{3}$ buffer layer/6H-SiC assuming Na coverage of 0.63 single layers. (b) The Brillouin zone of
1$\times$1 clean graphene (red hexagon) and that of the $\sqrt{3}\times\sqrt{3}$ buffer layer (blue hexagon). The
second Brillouin zone of the $\sqrt{3}\times\sqrt{3}$ phase shifted by 1.20 \AA$^{-1}$ is also shown by pale blue
hexagon. Calculated band structures (c) for the $\sqrt{3}\times\sqrt{3}$ buffer layer/6H-SiC system and (d) for the
single layer graphene/buffer layer/6H-SiC system. Note that the $\pi$ bands of the buffer layer near Fermi energy in
the vicinity of $\Gamma$ point appeared to be partially occupied in (c) for the buffer layer while they are empty in
(d) for the graphene layer.}
\end{figure}

Interestingly, the $\pi$ band quickly fades away with the development of the new band as Na dose increases on the
single layer graphene, which is more apparent in Fig.~\ref{fig:fig4}(a) where the momentum distribution curves (MDCs)
change with Na dose. The $\pi$ band becomes almost disappeared into the broad background, whereas the new band enhances
its spectral intensity showing the two peaks marked by arrows of the parabolic band centered at around 1.15 \AA~. The
$\pi$ band, however, behaves quite differently with Na for the bi-layer graphene mostly maintaining its intensity as
seen in Fig.~\ref{fig:fig4}(b). The rapid quenching of the $\pi$ band of the single layer graphene reflects a
destructive effect associated with the Na adatoms on the top graphene layer. The sustaining intensity of the $\pi$ band
from the bi-layer graphene in (f-j) may thus be attributed to the contribution from the second layer graphene
overcoming the loss from the top graphene layer. This idea seems to be supported by the change of band width as a
function of Na dose shown in inset of Fig.~\ref{fig:fig4}(b). The band width decreases initially down to the value of a
single layer graphene and then begins to increase with further dose of Na. We will discuss later the possible causes of
such changes of the $\pi$ band in Fig.~\ref{fig:fig2}.

We also performed {\em ab initio} pseudopotential total energy calculations with plane-wave basis set.\cite{Cohen} The
exchange-correlation of electrons was treated within the generalized gradient approximation (GGA) as implemented by
Perdew-Berke-Enzelhof.\cite{Perdew} The cutoff energy for the expansion of wave-functions and potentials in the
plane-wave basis was chosen to be 400 eV, and the Brillouin zone sampling was done with the Monkhorst-Pack special
k-point method with a grid of 5$\times$5$\times$1 for the $\sqrt{3}\times\sqrt{3}$ buffer layer phase. We used the
projector augmented wave pseudopotentials as provided by the software package (Vienna Ab-initio Simulation
Package).\cite{Kresse} The atomic relaxation was carried out until the Helmann-Feynman forces were less than 0.02
eV/\AA. The vacuum layer in the supercell used in our calculations is set to be 8 \AA, which is enough to minimize the
artificial interlayer interaction.

Figure~\ref{fig:fig5} shows the plausible binding site of Na atoms on the $\sqrt{3}\times\sqrt{3}$ buffer layer in (a)
and the Brillouin zones of the 1$\times$1 of graphite together with those of the buffer layer in (b). With the Na
coverage of 0.63 ML, the adsorption energy E$_{ad}$ calculated for the most favored site turns out to be 0.86 eV,
compared to the next favored hollow site of 0.72 eV. All hollow sites of graphene appear to have E$_{ad}$=0.73 eV. We
note that the SiC surface is not completely covered with either single- or bi-layer graphenes.\cite{Ohta08} We find the
most stable bonding configuration with a maximum adsorption energy E$_{ad}$=1.03 eV when Na atoms form a single
crystalline layer corresponding to the (110) plane of BCC Na as also seen in the scanning tunneling microscopy
measurements of Na islands on graphite.\cite{Brei01}

In Figure~\ref{fig:fig5}, we also present our band structures calculated with Na-added on the buffer layer in (c) and
on the graphene layer in (d). One immediately finds the three partially filled bands near the $\Gamma$ point in (c)
originating the upper two from the $\pi$ electrons in the buffer layer and the lower one from the Na adatoms. Since the
buffer-layer $\pi$ bands are completely empty without Na adatoms\cite{Kim07}, the $\pi$ bands of the buffer layer are
rigidly shifted down by charge transfer from Na adatoms to the $\pi$ bands. The nearly empty upper $\pi$ bands in (d)
then indicate that the charge transfer from Na occurs only to the top graphene layer with the underlying buffer layer
almost intact. As depicted in Figure~\ref{fig:fig5} (b) since the separation between the first ($\Gamma_0$) and second
($\Gamma_1$) Brillouin zones of the buffer layer is $\Delta$k=1.20 \AA$^{-1}$ much shorter than $\Delta$k=2.10
\AA$^{-1}$ of the Na (110) plane, the new parabolic band observed in Figure~\ref{fig:fig2} must be the $\pi$ bands of
the buffer layer partially filled by electrons from Na adatoms. The effective masses of the two buffer-layer $\pi$
bands are calculated to be 0.63 $m_0$ and 1.35 $m_0$, respectively. We note that the effective mass of the upper one is
quite close to the observed value of 0.58$\pm$0.07 $m_0$. The relatively weak parabolic bands seen from the bilayer
graphene in Figure~\ref{fig:fig5} (f-j) may be caused by the reduced area of the exposed buffer layer compared to that
of the single layer graphene.

\begin{figure}[t]
\includegraphics{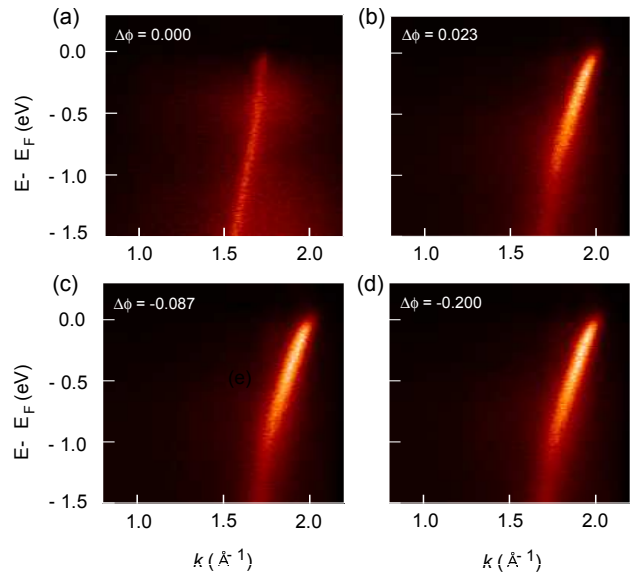}
\caption{\label{fig:fig6} (Color online) (a)-(d) The corresponding changes as in (a)-(e) of Fig. 2 with the sample at
RT. The changes are quite different from those in Fig. 2 showing the $\pi$ band rigidly shifted toward the higher
binding energy with enhanced spectral intensity. One also notes the absence of a new parabolic band near Fermi energy
in the vicinity of $\Gamma$.}
\end{figure}

We now consider why the linear $\pi$ band of graphene deteriorates with Na adsorption. One may think of two plausible
causes ; 1) due to the random scattering of photoelectrons from the disordered Na adatoms or 2) due to the reduced
mean-free path of charge carriers by enhanced back-scatterings of charge carriers from a Na-induced periodic potential.
The first cause, however, can be easily ruled out by our data in Fig.~\ref{fig:fig6} showing the corresponding behavior
of the $\pi$ band of graphene at RT. Here one immediately notices neither a new parabolic band nor the deterioration of
the $\pi$ band with increasing Na. Instead, the $\pi$ band becomes stronger in intensity and shifts toward the higher
binding energy by about 0.5 eV. Interestingly we find that the strong $\pi$ band at RT begins to deteriorate upon
cooling the sample eventually reproducing the low-temperature features in Fig.~\ref{fig:fig2}. The change in the $\pi$
band of graphene with Na appears to be reversible in temperature. We thus attribute the second cause to the Na-induced
weakening of the $\pi$ band of graphene assuming that Na adatoms form crystalline islands at low temperature giving
rise to a certain form of a periodic potential.

If Na crystalizes at 85 K in the form of a layer corresponding to the (110) plane of BCC Na crystal as the Na islands
formed on graphite,\cite{Brei01} the periodicity L of the Na-induced potential may be similar to a$_0$ of the graphene
lattice constant. Such a potential may severely modify the band structure in a way to reduce the mean-free path of
charge carriers within graphene by enhanced back-scatterings of charge carriers as noted
earlier.\cite{Park08,Ando1,Ando2,McEuen99} The formation of such crystalline Na islands at 85 K is indeed highly likely
considering the significantly reduced hopping rate between the hollow binding sites ($\nu=\nu_0 e^{-E_d/k_BT}$, where
$\nu_0$ is an attempt frequency and $k_B$ is the Boltzmann constant), which is about 3.65$\times$10$^5$ times smaller
than that at RT. Here we have used the diffusion barrier of 0.13 eV between the most stable hollow sites for Na on
graphene. The stronger and shifted $\pi$ band at RT suggests that the $\pi$ band is shifted down rigidly towards the
higher binding energy side revealing the highly occupied band near Fermi level as also observed when K is
absorbed.\cite{Aaron} On the other hand, the increased hopping rate of Na at RT drives delocalization of charge
carriers\cite{Ahn} and reduces the donation of charges to the buffer layer leaving the $\pi$ band of the buffer layer
empty as we find Fig.~\ref{fig:fig6}.

\section{Conclusion}
By adsorbing Na on graphene at two different temperatures (RT and 85 K) we observe that the band structure of graphene
formed on SiC substrate becomes quite sensitively modified by the Na-induced periodic potential especially when the
periodicity of the potential is comparable to the lattice constant of graphene. The frozen $\pi$ band of graphene
quickly weakens in intensity and broadens in line-width with Na added at low temperature while it becomes stronger but
is solidly shifted towards the higher binding energy with Na at RT. A new parabolic band observed only at 85 K by
adding Na is found to be the $\pi$ band of the underlying buffer layer partially filled by the charges from Na. Such an
extremely sensitive nature of Dirac fermions in graphene to external impurities as well as temperature demands a more
careful study of transport properties of graphene for future practical applications in developing graphene based
electronic devices.

\acknowledgments This work was supported by the Korea Science and Engineering Foundation (KOSEF) grant funded by the
Korea government (MEST) by R01-2008-000-20020-0, and also in part by NCRC grant R15-2008-006-01001-0.


\end{document}